# Physical, optical and nonlinear properties of InS single crystal

Pallavi Kushwaha[1], Anuradha Patra[1]*, E. Anjali[2], Harshad Surdi[1], C. Gurada[3], S. Ramakrishnan[1], S. S. Prabhu[1], Venu Gopal Achanta[1] and A. Thamizhavel[1]

[1]DCMP&MS, Tata Institute of Fundamental Research, Homi Bhabha Road, Mumbai 400 005 India
[2]National Institute of Technology Calicut, 673 601 India
[3]Department of Physics, University of Mumbai, Mumbai 400 098 India
*anuradha.p@tifr.res.in

**ABSTRACT**

Indium Sulphide (InS) single crystals are successfully grown by In flux. Single crystal X-ray diffraction shows orthorhombic structure of *Pnnm* space group. Ellipsometry measurements performed on the (010) oriented crystal exhibit low anisotropy in the 300-1000 nm wavelength range and consequently negligible THz emission is observed. Optical band gap of 2.09eV is deduced from linear optical measurements. Nonlinear optical properties are studied by single beam Z-scan measurements at 800 nm, where two-photon absorption is present. Nonlinear refractive index and absorption coefficient are estimated to be $n_2 = 2.3 \times 10^{-11} cm^2/W$ and $\beta = 62.4 cm/GW$, respectively for excitation intensity of 0.32 GW/cm². The origin of nonlinearity in InS crystal is accounted to be due to the third-order anharmonic motion of the bound electrons.

**Keywords: Indium Sulphide, Single crystal, THz, Z-scan**

# 1. Introduction

Indium sulfide is a semiconducting compound belonging to III-VI family. Some of the prominent members in this family are layered crystals of GaS, GaSe and InSe. Indium sulfide, like its counterparts, coordinates with its constituents in the tetrahedral geometry. However, the two S atoms and one In atom are in one plane and uniquely, the third S atom is in the neighboring layer. This makes the crystal structure of indium sulphide as a three dimensional network. Indium sulphide has strong anisotropic features in its crystal structure and crystallizes in three crystallographic configurations, namely, α, β, and γ-$In_2S_3$. The most stable one is $β$-$In_2S_3$ which has a tetragonal structure [1]. Being a wide band gap semiconductor [2], thin films of $In_2S_3$ are investigated for optoelectronics and photovoltaic applications [3]. Thin films of $In_2S_3$ are also investigated as a solar energy absorbing material [4], and may be a suitable alternative to Cadmium sulfide based solar cells which include the toxic Cadmium [5]. Consequently, studies on thin films of $In_2S_3$ were reported [1-6]. However, reports on single crystal InS are scarce [7-14]. partly owing to the difficulty in growing single phase samples. The phase diagram of In and S clearly indicates that the equiatomic composition of In and S is incongruently melting with a dissociation temperature at 600 C [15]. Electroreflectance measurements on single crystal InS [7], grown from In melt, is reported in the energy range from 2-4 eV. The authors have determined the direct transition edges to be 2.45 eV, 2.57 eV and 2.58 eV at temperatures of 290 K, 93 K and 10 K respectively. Nishino and Hamakawa have performed electrical and optical measurements on an InS single crystal synthesized using solution growth method from In melt [8]. They reported n-type conductivity of InS at room temperature and have estimated room temperature indirect and direct band gaps to be 1.9 eV and 2.44 eV respectively. Polymorphism of InS at high pressure has been reported by Kabalkina *et al* [9]. It has been found that at high pressure (7.5 GPa /6 GPa at and 293 K/ 573) and isothermal conditions, the crystal of InS undergoes transition from orthorhombic to tetragonal structure [9]. High pressure Raman and transport measurement were also performed on InS single crystal by Faradzhev *et al.* [10] and Takarabe *et al.* [11] respectively. Studies on the electrical resistance of single crystal InS at pressures upto 150 kbar, using uniaxial split-sphere apparatus, indicate possible insulator- metal transition at high pressure. The results have been attributed to the change in the band structure due to increase in pressure. In another interesting work by Takarabe *et al.* [12] showed that the difference in the crystal structures between InS and other III-VI layered semiconductors counterparts can be explained in terms of effective changes or ionicities. The authors have studied infrared reflection spectra of single crystal InS in the wave number range between 50 and 900 $cm^{-1}$ for each polarization in *c*-plane [12]. Gasanly and Aydinli [13] have reported low temperature photoluminescence spectra of InS single crystal excited by the 476.5 nm argon ion laser and have observed three photoluminescence bands centered at 605 nm (A-band), 626 nm (B-band) an 820 nm (C-band). Later, Qasrawi and Gasanly [14] have retrieved information about the localized levels in the forbidden gap of InS crystal from the results of dark electrical conductivity, space-charge limited current and photoconductivity measurements in the temperature range of 10-350 K. Although, few reports on the structural, electrical and optical studies on InS are documented in literature, to our knowledge, there is no report on the nonlinear optical properties of InS single crystals. Z-scan is an established technique to measure the optical nonlinearities in a material [16,17]. Briefly, in Z-scan, the sample is irradiated by high intense laser beam of variable intensity and the transmitted intensity pattern is mapped in sample's transverse plane. The technique, despite being simple, simultaneously measures the magnitude as well as the sign of nonlinear absorption and refraction with high sensitivity. Depending on the sign of nonlinear refraction, the modification in the refractive index results in self focusing or defocusing effects, while that in the absorption coefficient leads to induced transmission or induced absorption effects. In the present work, we synthesized single crystal InS by flux method and present their structural, linear and nonlinear optical properties.

In the nonlinear optical regime, the refractive index and absorption coefficient are intensity dependent and leads to wave mixing, harmonic generation etc [16]. Briefly, in Z-scan, the sample is irradiated by high intense laser beam of variable intensity and the transmitted intensity pattern is mapped in sample's transverse plane. The technique, despite being simple, simultaneously measures the magnitude as well as the sign of nonlinear absorption and refraction with high sensitivity. Depending on the sign of nonlinear refraction, the modification in the refractive index results in self focusing or defocusing effects, while that in the absorption coefficient leads to induced transmission or induced absorption effects. In the present work, we synthesized single crystal InS by flux method and present their structural, linear and nonlinear optical properties.

## 2. Sample preparation and experimental details

InS single crystals were prepared by flux method starting with high purity constituent elements of InS in *In* flux was used. The sample (InS) to flux (In) ratio is taken as 1:19 weight percentage. Required Sulfur powder was repetitively sandwiched and rolled between Indium plates to reduce the Sulfur evaporation. This mixture was kept in alumina crucible and then sealed in a quartz ampule in $10^{-6}$ Torr. The sealed ampule was placed in the furnace and heated up to 1050 C with the rate of 35 C/hrs. The temperature was kept for a day. The crystallization was made by cooling the system to 550 C at a rate of 5 C/hrs. Most of the excess Indium was separated by centrifuging. However, a small amount was stuck at the edge of the sample. As grown crystals were in platelet form with clean natural faces that are like cleaved ones and having red color. EDAX measurements showed that the as grown red crystals are having the equiatomic In:S stoichmetry of 1:1. The crystals are of centimeter size and thickness varied from few tens of μm to hundreds of μm (photograph of as grown crystal is shown in the inset of Fig.1).

## 3. Result and discussion

3.1 Crystal structure analysis

There is very little literature on InS system and even the existing ones reported different crystal structure for the equiatomic InS. Thus, it is not clear if InS system is monoclinic or orthorhombic [9,18]. To clear this issue, single crystal x-ray analysis is performed on the grown crystal. A small piece of 0.2 mm× 0.2 mm× 0.05 mm dimension was cut from the big crystal after a thorough investigation under the optical microscope used for single crystal X-ray measurement. X-ray intensity data for the Indium and Sulfur were collected at 275 K on a Bruker Axs Kappa APEX2 diffractometer equipped with graphite monochromated Mo (Ka) radiation (λ = 0.71073 Å). The software programs used for the data collection, cell refinement and data reduction are APEX2 , SAINT-Plus and XPREP respectively [19]. The automatic cell determination routine, with 36 frames at three different orientations of the detector was employed to collect reflections for unit cell determination. The lattice parameters are found to be a = 4.4506(2) Å; b = 10.6503(4) Å; and c = 3.9455(2) Å and α = β = γ = $90^0$. Intensity data were collected for one hemisphere (h= ±5; k=±13; l=±4). A total of 4676 reflections with 3.83 ≤ θ ≤ 26.36 were recorded and multi-scan absorption correction (SADABS) was applied. After absorption correction, systematic absences showed the existence of orthorhombic symmetry with *Pnnm* space group. Two hundred and twenty two unique data were obtained after merging of equivalent reflections.

The structure was solved by direct methods (SHELXS-97), and the atoms were refined anisotropically using full matrix least squares on $|F|^2$ with all unique reflections (SHELXL-97). The atom positions are unambiguously identified. Table 1 shows the fractional coordinates, site occupancies, and thermal parameters of the atoms in the asymmetric unit. The structure is refined with a residual factor of 0.0295 and goodness of fit value of 1.284. Figure 1 shows the atomic arrangement of In and S atom in unit cell generated by using parameter listed in table 1. Inset to the Figure 1 shows the photograph of as grown crystal and, the Laue diffraction on these surfaces confirmed that this plane corresponds to the (010) plane. Figure 2 shows the observed and generated powder x-ray diffraction for InS. For clarity only high intense peaks are indexed. All observed peaks match well with generated pattern except peaks marked by * corresponds to elemental In which is used as the flux to grow the crystals. As shown by the image (a) in the inset of Fig.2, SEM images clearly shows excess indium sticking to the sample surface. However cross section SEM image shows pure InS in bulk as shown in image (b). From XRD data, one can notice that the highest intense peaks are different in observed and generated data. It may be due to the contribution of the little extra indium and also due to the fact that the grown InS crystals are malleable in nature, which preserves the crystal from becoming powder. From Laue diffraction, we find that the crystal surface corresponds to (010) plane so in powder pattern peak analogous to (010) like (040) and (080) shows higher intensity compared to other peaks because of preferred orientation.

**TABLE I**. *Fractional atomic coordinates and thermal parameters for InS. Atomic coordinates ( $\times 10^4$ ) and equivalent isotropic displacement parameters ($A^2 \times 10^3$ ) for InS. U(eq) is defined as one third of the trace of the orthogonalized Uij tensor.*

| Atom | x | y | z | site occ. | $U_{eq}$ | $U_{11}$ | $U_{22}$ | $U_{33}$ | $U_{23}$ | $U_{13}$ | $U_{12}$ |
|---|---|---|---|---|---|---|---|---|---|---|---|
| In(1) | 1221(1) | 6191(1) | 5000 | 1 | 12(1) | 16(1) | 9(1) | 12(1) | 0 | 0 | -2(1) |
| S(1) | -199(4) | 8472(1) | 5000 | 1 | 11(1) | 13(1) | 9(1) | 10(1) | 0 | 0 | 0 |

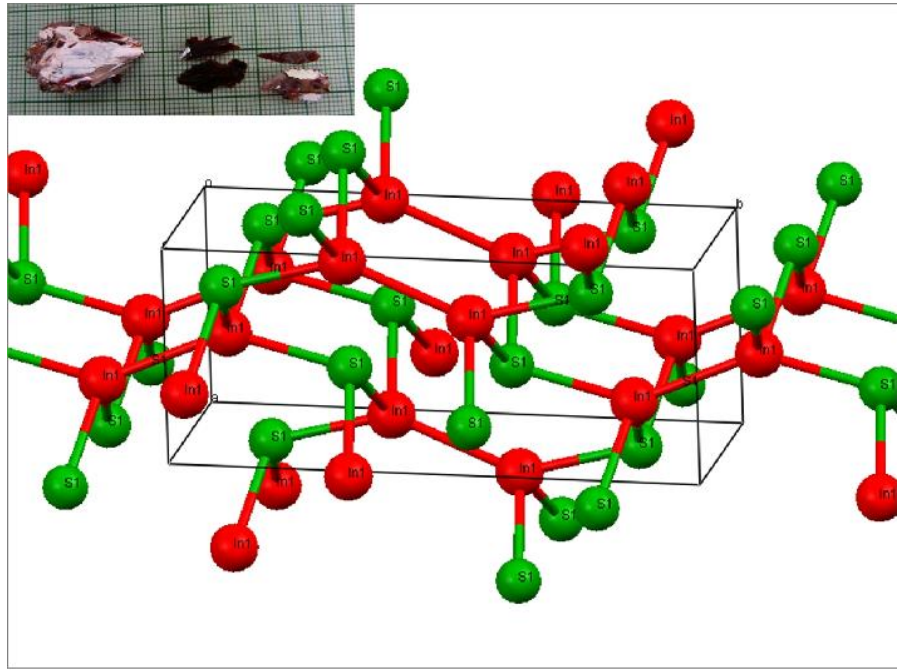

*FIG. 1. Atomic arrangement of In and S atom in unit cell generated by using parameter listed in table 1. Inset shows the photograph of as grown crystal.*

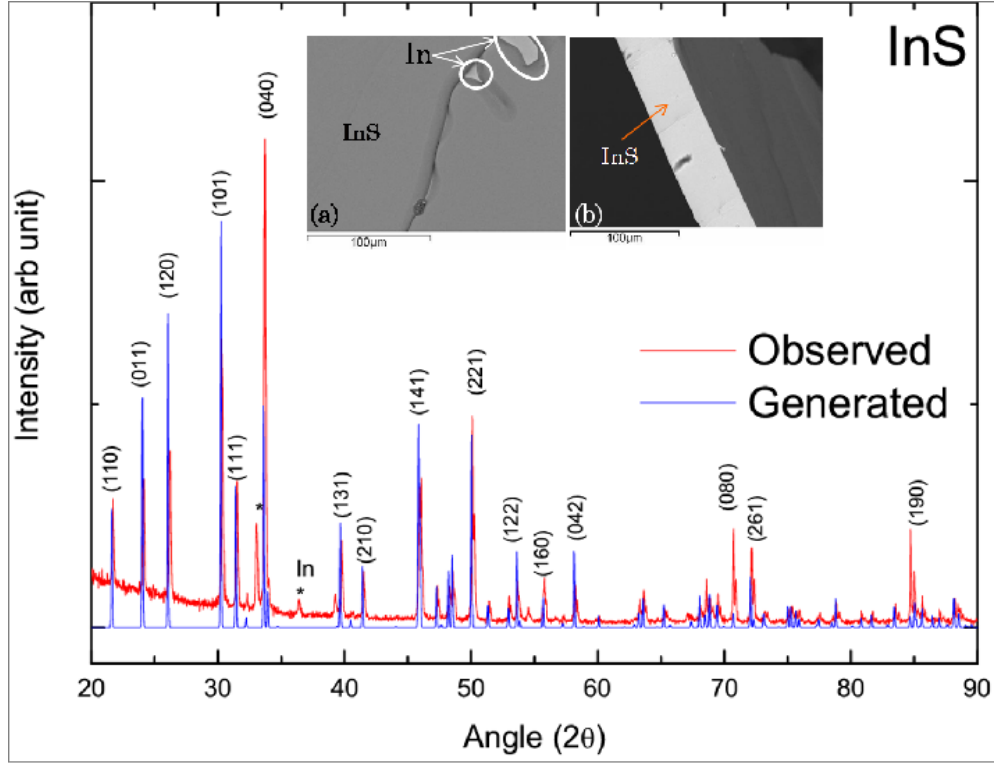

*FIG. 2. Observed powder x-ray diffraction (red line/upper pattern) pattern and generated pattern (lower pattern) using single crystal x-ray data of InS sample. All peaks are indexed. Peak marked by * are corresponds to elemental Indium. Inset shows the SEM images on (a) sample surface and (b) cross section area. Encircled area shows excess indium on sample surface.*

3.2 Linear and nonlinear optical studies on InS single crystal

The linear transmittance and reflectance spectra of single crystal InS were recorded at room temperature in JASCO V-670 spectrophotometer in 200 nm - 2500 nm wavelength range and are displayed in Figs.3 a and b. From the figures it is clear that the single crystal InS has low transmission and reflection. The variation of real and imaginary parts of the complex refractive index computed from the ellipsometry data are shown in Fig. 3c. All the measurements were performed on a 50 μm thick film. The absorption coefficient, $\alpha$ was calculated from the relation, $\alpha = 4\pi k / \lambda$, where $k$ is the extinction coefficient, which is the imaginary part of the complex refractive index and λ is the wavelength. Optical band gap of InS thin film was calculated from the plot of $(\alpha h\nu)^2$ against photon energy $(h\nu)$ and is shown in Fig. 3d as open circles. From the intercept of the straight line fit to the linear portion of the curve ($\chi^2$ minimization fit shown in Fig. 3d), band gap value was found to be 2.09 eV. There is considerable deviation in reported band gap values. Single crystal β-$In_2S_3$ is reported to have band gap in the range of 2 eV to 2.4 eV, whereas polycrystalline β-$In_2S_3$ have values ranging form 2 to 3.7 eV [6]. The possible reasons behind such variations are different growth conditions and stoichiometry differences. In fact, the freedom to tailor the band gap of $In_xS_y$ by changing the stoichiometry has been the prime factor for its potential application in solar cells [6]. It is also important to note that $\beta$-$In_2S_3$ crystallizes in tertragonal structure whereas, the structure of stoichimetric InS single crystal in the present study has been recognized to be orthorhombic. No prior information about THz transmission below 3 THz [12] in InS single crystal has been reported so far. Ten femtosecond laser of wavelength 800 nm and repetition frequency of 80 MHz was focused onto InS crystal to study the THz transmission in the 0.2 to 3.0 THz region. Measured transmission (figure not shown) showed very weak transmission. The low THz transmission could be due to high conductivity of InS crystal.

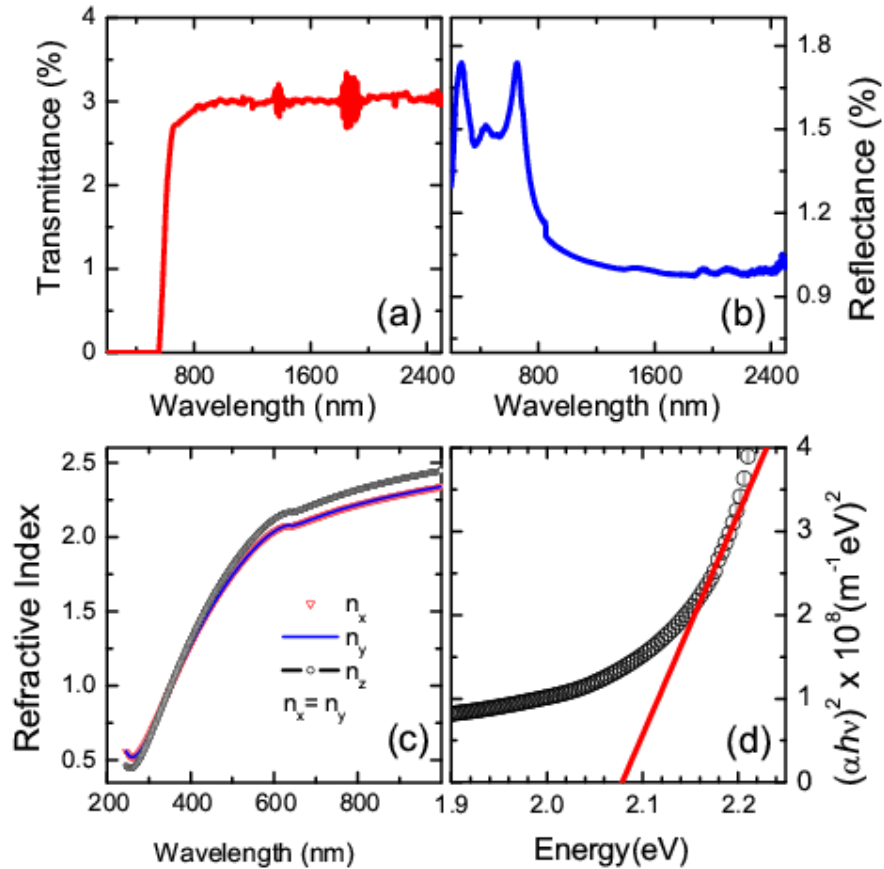

**FIG. 3** *Linear optical studies on single crystal InS (a) Transmittance spectrum (b) Reflectance spectrum (c) Refractive index profile (d) Variation of $(\alpha h\nu)^2$ vs photon energy, $h\nu$. The line shows the fit.*

The nonlinear optical absorption characteristics of the single crystal InS have been investigated by performing Z-scan measurements by using femtosecond laser pulses. In the nonlinear regime, the intensity (*I*) dependent refractive index (*n(I)*) and absorption coefficient (*α(I)*) are given by [16],

$$n(I) = n_0 + n_2 I$$
$$\alpha(I) = \alpha_0 + \beta I$$

where $n_0$ and $\alpha_0$ are the linear refractive index and absorption coefficient, respectively. $n_2$ and $\beta$ are the nonlinear refractive index and absorption, respectively. We performed open aperture (OA) Z-scan where the irradiance at the sample is varied by translating the sample. This mode of operation is sensitive to nonlinear absorption (β), as any deviation in the transmitted intensity must be due to multi-photon absorption. Placing a limiting aperture past the focus of the lens and before the detector, a typical arrangement for closed aperture

(CA) mode, however, leads to modulation of transmitted intensity in an anti-symmetric way with respect to the focal point. The detector, due to the presence of an aperture is now sensitive to nonlinear refraction in the sample. Any focusing or defocusing of the sample may manifest itself as either beam broadening or narrowing in the far field and this leads to a shaped dispersion curve from which nonlinear refraction (α) is easily estimated. The Z-scan measurements have been performed by launching femtosecond laser of wavelength 800 nm with repetition frequency 80 MHz onto the InS crystal. The incident beam is focused by a lens of focal length 12 cm to provide the beam waist radius of $\omega_0 = 50 \mu m$ and the diffraction length ($z_0$) of 10 mm. Since the sample thickness is much smaller than the diffraction length, thin film approximation has been used to simulate the Z-scan data. The measurements were recorded at the peak irradiance of 0.32 GW/cm². InS ($E_g$ = 2.09 eV) is a two-photon absorber at 800 nm. In the limit of two-photon absorption, the normalized change in the OA transmittance is given by the expression [20]

$$T = 1 - \frac{\beta I_0 L_{eff}}{\left(1+\left(\frac{z}{z_0}\right)^2\right)2^{\frac{3}{2}}}$$

where, the effective thickness of the sample denoted as $L_{eff}$ is defined by the expression,

$L_{eff} = \frac{1-e^{-\alpha_0 l}}{\alpha_0}$, with $\alpha_0, l$ being the linear absorption coefficient and sample thickness, respectively. $I_0$ is the on-axis power density of the laser beam at the waist. Typical OA scan of InS film is displayed in Fig.4a with the fit shown by a continuous line. The nonlinear absorption has been estimated to be $\beta = 62.4 cm/GW$.

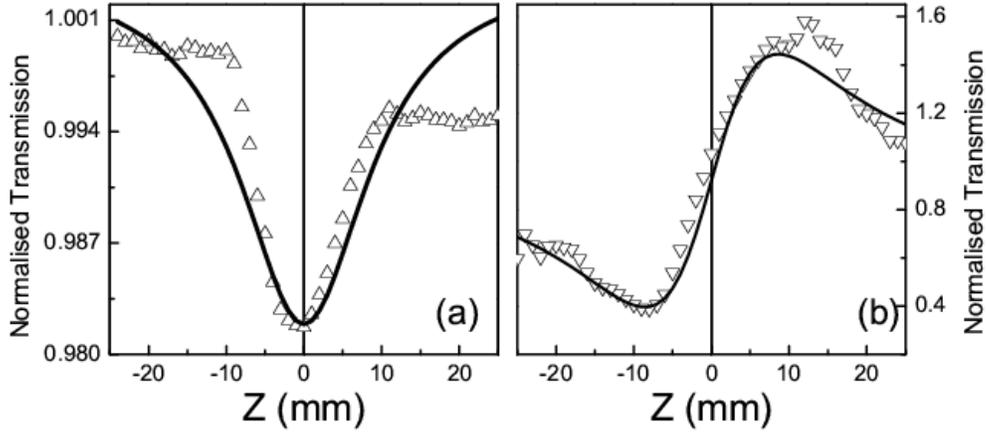

***FIG.4*** *Nonlinear optical studies on single crystal InS (a) Open aperture and (b) Closed aperture Z-scan. The symbols represent experimental data and continuous line through the data points represents fit.*

For an optical beam at wavelength $\lambda$, the light induced nonlinear refractive index $n_2$ is related to the nonlinear phase shift $\Delta\Phi_0$ by the relation, $\Delta\Phi_0 = \frac{2\pi}{\lambda} n_2 I_0 L_{eff}$. The sign of $n_2$ is thus determined by $\Delta\Phi_0$. An important feature of CA Z-scan is that the sign of $\Delta\Phi_0$ and $n_2$ as a consequence, is determined from the relative position of the peak and valley. Typical valley-peak configuration, i.e, pre-focus valley followed by post-focus peak in the CA trace is the signature of self-focusing action of the sample whereas, the peak-valley transmittance profile is the manifestation of self-defocusing in the sample [20]. Fig. 4b represents the CA scan of the InS crystal. The experimental data were fitted with the following equation [20] to derive nonlinear refractive index $n_2$,

$$T = 1 - \frac{4\Delta\Phi_0 \left(\frac{z}{z_0}\right)}{\left(1+\left(\frac{z}{z_0}\right)^2\right)\left(9+\left(\frac{z}{z_0}\right)^2\right)}$$

In Fig. 4b the symbols represent the experimental data and solid line shows the fit. From the valley-peak configuration, it is clear that the InS film acts as self-focusing lens of variable focal length. The best fit provides $n_2 = 2.3 \times 10^{-11} cm^2/W$. The linear refractive index, computed from ellipsometry measurement is 2.3 at 800 nm. From the difference between the normalized peak and valley transmittances ($\Delta T_{p-v}$) in CA scan which is estimated close to ~1.7 $z_0$, it is evident that cubic nonlinearity prevails in the present case. Moreover, at irradiance of 0.32 GW/cm² it is expected that the changes in the index of refraction is due to the third-order anharmonic motion of the bound electrons. The free-carrier contribution to refraction becomes significant for higher irradiance level [21]. It has also been shown [22] that for wavelengths closer to the energy gap, electronic Raman effect and ac Stark effect play a dominant role. These effects may also be absent or negligible at 800 nm for InS film.

## 4. Conclusion

Structural, optical and nonlinear optical properties of large size single crystal InS have been studied. The structure of single crystal InS, determined from single crystal X-ray diffraction, has been recognized to be orthorhombic of *Pnnm* space group. InS crystal has been found to have a direct band gap of 2.09 eV, low transmittance and reflectance in the UV-VIS-NIR wavelength range, exhibit low THz transmission and shows anisotropy in refractive index profile. Nonlinear optical measurements by Z-scan technique exhibits nonlinear absorption and self focusing effects. At irradiance of 0.32 GW/cm², the nonlinear refractive index and absorption coefficient at 800 nm are estimated to be $n_2 = 2.3 \times 10^{-11} cm^2/W$ and $\beta = 62.4 cm/GW$ respectively.

**Acknowledgement**

We thank Dr. Babu Varghese, from Sophisticated Analytical Instrument Facility, Indian Institute of Technology Madras, Chennai for single crystal X-ray diffraction measurement.